\documentclass[pdflatex,sn-mathphys-num]{sn-jnl}

\usepackage{graphicx}%
\usepackage{multirow}%
\usepackage{amsmath,amssymb,amsfonts}%
\usepackage{amsthm}%
\usepackage{mathrsfs}%
\usepackage[title]{appendix}%
\usepackage{xcolor}%
\usepackage{textcomp}%
\usepackage{manyfoot}%
\usepackage{booktabs}%
\usepackage{algorithm}%
\usepackage{algorithmicx}%
\usepackage{algpseudocode}%
\usepackage{listings}%
\usepackage{amsbsy}
\usepackage{lipsum}

\usepackage{pgfplots}
\usepackage{subfig}
\usepackage{graphicx}
\usepackage{siunitx}  
\usepackage{adjustbox}
\usepackage{changepage}

\theoremstyle{thmstyleone}%
\newtheorem{theorem}{Theorem}
\newtheorem{proposition}[theorem]{Proposition}%

\theoremstyle{thmstyletwo}%

\theoremstyle{thmstylethree}%
\newtheorem{definition}{Definition}%
\DeclareMathOperator{\vect}{vec}
\DeclareMathOperator{\tr}{tr}
\begin{document}

\title[Article Title]{Optimal design of experiments for functional linear models with dynamic factors}


\author*[1,2]{ \sur{Caterina May}}\email{caterina.may@kcl.ac.uk}

\author[1]{ \sur{Theodoros Ladas}}\email{theodoros.ladas@kcl.ac.uk}

\author[1]{ \sur{Davide Pigoli}}\email{davide.pigoli@kcl.ac.uk}

\author[1]{ \sur{Kalliopi Mylona}}\email{kalliopi.mylona@kcl.ac.uk}

\affil*[1]{\orgdiv{Department of Mathematics}, \orgname{King's College London}, \orgaddress{\street{Strand Campus}, \city{London}, \postcode{WC2R 2LS}, \country{UK}}}
\affil[2]{\orgdiv{Dipartimento DiSEI}, \orgname{Università del Piemonte Orientale}, \orgaddress{\street{V. Ettore Perrone 18}, \city{Novara}, \postcode{28100},  \country{Italy}}}


\abstract 
   {In this work we build optimal experimental designs for precise estimation of the functional coefficient of a function-on-function linear regression model where both the response and the factors are continuous functions of time. After obtaining the variance-covariance matrix of the estimator of the functional coefficient which minimizes the integrated sum of square of errors, we extend the classical definition of optimal design to this estimator, and we provide the expression of the A-optimal and of the D-optimal designs. Examples of optimal designs for dynamic experimental factors are then computed through a suitable algorithm, and we discuss different scenarios in terms of the set of basis functions used for their representation. Finally, we present an example with simulated data to illustrate the feasibility of our methodology. 

   }




\maketitle

\section{Introduction}\label{sec1}

In many applications data are collected as curves over time, and it is therefore natural to consider statistical models with curves (or functions) as response and as covariates. This kind of problems has been considered for example in biomedical sciences (\cite{shi2012mixed}),  genetics (\cite{muller2008inferring}), pharmaceutical sciences (\cite{rahman2019functional}) and transport engineering (\cite{zhang2018functional}), to mention just a few. However, most of these works have been concerned with observational data and little attention has been devoted to experimental design for settings where both the covariates and the response are functional in nature. An example of a laboratory experiment that fits in this framework and motivated this research is the one described in \cite{Pigoli23}, in the context of forensic entomology. Entomologist are interested in studying the relationship between larval growth (and growth rate) and temperature, but all these quantities vary in time. In a laboratory setting, the current practice is to keep the incubator temperature constant for each run, and then having various runs at different temperatures. However, this is a choice based only on practicality, since it doesn't require the operator to change the incubator temperature during the run. A natural question is if there are better experimental designs to explore the relationship between the growth rate curve and the temperature curve. This is an example of the kind of the scenarios we wish to address in this work.     

We consider a framework where both the experimental factors and the responses have a functional nature. In practice, the data analysis always start from raw measurements and a pre-processing smoothing step is required to represent observations as functions. We refer to \cite{Ramsay2005} for an extensive discussion of the smoothing procedures that can be used to reconstruct the functional data; the starting point of our work will be responses and factors that are already represented as functions. Also, we restrict ourselves to the case of \emph{densely observed} functional data \cite{gertheiss2024functional}, where smoothing can indeed be treated as a pre-processing step, while we will not discuss the case of \emph{sparsely observed} functional data, where the location of the measurements can pose in itself an experimental design issue.

In terms of statistical models, in this work we will consider function-on-function linear models (see, e.g. \cite{Ramsay2005} or \cite{HK}). To the best of our knowledge, statistical inference on these models has not received much attention. The original contribution of this paper is therefore twofold. First, we present new inferential results regarding the estimation of the functional coefficient. Second, we use these results to construct optimal experimental designs with dynamic factors, for precise estimation of the functional coefficient. This will expand on the current state-of-the-art on experimental design for functional data,  where optimal designs for experiments for scalar-on-function linear models (i.e. models with functional covariates but scalar response) has been considered by \cite{michaelidesArx} and for function-on-scalar linear models has been discussed in 
\cite{aletti2014optimal} and \cite{AMT_Moda11}. 

The paper is organized as follows. Section \ref{sec:model} contains the description of the model. In Section \ref{sect:estim} the estimator of the functional coefficient of multiple factors is derived, and its properties are proved. Section \ref{sec:OptDes} is devoted to the construction of exact optimal experimental design on the base of the inference results previously obtained. In addition, Section \ref{sec:ex} illustrates and discusses some concrete examples of optimal designs, obtained with an optimization algorithm. Finally, a simple example where the coefficient is estimated from simulated data, useful for practitioners as an illustration of our methodology, is provided in Section \ref{sec:sim_ex}. A final section with future developments concludes this work.

\section{Model and notation}\label{sec:model}
Throughout this paper, we consider the following function-on-function linear model:
\begin{equation}\label{model1}
y_n(t)=\beta_0(t)+\sum_{i=1}^p\int_0^{\cal T} \beta_i(s,t)x_{ni}(s)ds + \varepsilon_n(t), \quad n=1, ...,N, \quad p\ge 1
    \end{equation}
 where $\varepsilon_1(t), \hdots, \varepsilon_N(t)$ are i.i.d. zero-mean stochastic processes  (not necessarily Gaussian). In 
 matrix notation, model \eqref{model1} can be rewritten as

 \begin{equation}\label{model1mat}
\mathbf{y}(t)=\int_0^{\cal T} \mathcal{X}(s) {\pmb{\beta}}(s,t)ds + {\pmb{\varepsilon}}(t), 
    \end{equation}

    where
    \begin{itemize}
        \item[] $\mathbf{y}(t)=(y_1(t), \hdots, y_N(t))^T$ is the vector of responses,
        \item[] $\pmb{\varepsilon}(t)=(\varepsilon_1(t), \hdots, \varepsilon_N(t))^T$ is the vector of errors,
        \item[] $\pmb{\beta}(s,t)=(\beta_0(t), \beta_1(s,t), \hdots, \beta_p(s,t))^T$ is the functional coefficient to be estimated,
        \item[] ${\cal X}(s)$ is the $N\times (p+1)$ model matrix:
         \begin{equation}\label{eq:Xmatrix}
        \mathcal{X}(s)=[\mathbf{1}_N \,|\,\mathrm{X}(s)],
        \end{equation}

    \item[] with $\mathrm{X}(s)$ being the $N\times p$ design matrix of the dynamic predictors:
        \begin{equation}\label{eq:Dmatrix}
        \mathrm{X}(s)= \begin{bmatrix}
            \mathbf{x}_1(s) & \mathbf{x}_2(s) & \hdots & \mathbf{x}_p(s)
        \end{bmatrix}
        =\begin{bmatrix}
            x_{11}(s) & \hdots & x_{1p}(s) \\
            x_{21}(s) & \hdots & x_{2p}(s) \\
            \vdots &   & \vdots \\
            x_{N1}(s) & \hdots & x_{Np}(s)  \\

        \end{bmatrix}
    \end{equation}
    \end{itemize}

We consider a basis $\{\theta_l(t), l\ge 1\}$ to expand the responses $y_n(t)$, and the errors $\varepsilon_n(t)$, for $n=1, ..., N$. 
We can then expand the intercept $\beta_0(t)$ and the kernels $\beta_i(s,t)$, for any $i=1, ..., p$, according to the basis $\{\theta_l(t), l\ge 1\}$
and to $p$ bases $\{\eta^1_k(s), k\ge 1\}$, ...,  $\{\eta^p_k(s), k\ge 1\}$:
\begin{equation*}
    \beta_0(t)=\sum_l b^0_l\, \theta_l(t),
\end{equation*}

\begin{equation*}
    \beta_i(s,t)= \sum_{k,l} b^i_{k,l}\, \eta^i_k(s)\, \theta_l(t).
\end{equation*}

Since, by  the Grand-Schmidt orthonormalization procedure, any basis can be transformed into an orthonormal basis, from now on we assume that the basis $\{\theta_l(t), l\ge 1\}$ is orthonormal, that is,
\begin{equation*}
    \langle \theta_h(t), \theta_k(t) \rangle _{L^2}= \int \theta_h(t) \theta_k(t) dt = \delta_{hk},
\end{equation*}
where is the Kronecker delta symbol: $\delta_{hk}=1$ if $h=k$ and $\delta_{hk}=0$ if $h\neq k$.

In order to estimate the parameters of the functional model \eqref{model1}, we follow the ideas contained in \cite{Ramsay2005} and in \cite{HK}; their fully functional model is here extended to multiple factors and, in Section \ref{sect:estim}, we derive explicitly the estimator for the parameters of our model.
Let us project the observed functional responses, the unknown coefficients and the dynamic factors, in the subspaces generated by the finite bases $\{\theta_l(t), 1 \le l \le L\}$ and $\{\eta^1_k(s), 1 \le k\le K^1\}$, ...,  $\{\eta^p_k(s), 1 \le k\le K^p\}$. Note that each factor $\mathbf{x}_i(s)$, for $i=1, ..., p$, and their respective coefficients $\beta_i(s,t)$ can be expanded, with respect to their first variable $s$, in a different basis with dimension $K^i$,  according to its features. Moreover, in some experimental situations, it could be useful to expand each factor with a basis $\{c^i_k(s), 1 \le k\le K_x^i\}$ which is not the same basis as the one used to expand its correspondent unknown coefficient (see Section \ref{sec:diff_bas}).

We have:
\begin{equation*}
    \beta_0^*(t)= {\mathbf{b}^0}^T \, \pmb{\theta}(t),
\end{equation*}

\begin{equation*}
    \beta_i^*(s,t)= \pmb{\eta}^i(s)^T \mathrm{B}^i \, \pmb{\theta}(t),
\end{equation*}

for any $i=1, ..., p$, where

\begin{equation*}
    {\mathbf{b}^0}=(b_1^0, \hdots, b_L^0)^T, 
\end{equation*}

\begin{equation*}
    \pmb{\theta}(t)=(\theta_1(t), \hdots, \theta_{L}(t))^T,
\end{equation*}

\begin{equation*}
    \pmb{\eta}^i(s)=(\eta^i_1(s), \hdots, \eta^i_{K^i}(s))^T, 
\end{equation*}

and

\begin{equation*}
    \mathrm{B}^i=[ b^i_{k,l}]
\end{equation*}
is a $K^i \times L$ matrix. 
Then the functional coefficient $\pmb{\beta}(s,t)$ 
becomes
\begin{equation}\label{eq:betastar}
   \pmb{\beta}^*(s,t)=(\beta_0^*(t), \beta_1^*(s,t), \hdots, \beta_p^*(s,t))^T = 
   \mathrm{H}(s)^T \, \mathrm{B}\, \pmb{\theta}(t)
\end{equation}
where $\mathrm{H}(s)^T$ is a $(p+1)\times (1+K^1+...+K^p)$ diagonal block matrix given by
\begin{equation*}
    \mathrm{H}(s)^T=\mbox{diag}(1, \pmb{\eta}^1(s)^T, \hdots,\, \pmb{\eta}^p(s)^T),
\end{equation*}
and $\mathrm{B}$ is an $(1+K^1+...+K^p) \times L$ super matrix of coefficients
\begin{equation*}
    \mathrm{B}=\begin{bmatrix}
        {\mathbf{b}^0}^T\\
        \mathrm{B}^1\\
        \vdots\\
        \mathrm{B}^p
         \end{bmatrix} 
\end{equation*}

\noindent  to be estimated.
Model \eqref{model1mat} becomes:
\begin{equation*}\label{model1mat2}
\mathbf{y}(t)= \int_0^{\cal T} \mathcal{X}(s) \pmb{\beta}^*(s,t)ds + {\pmb{\varepsilon}}(t)
    \end{equation*}
that is, putting the \eqref{eq:betastar} into it,
\begin{equation}\label{model1mat3}
\mathbf{y}(t)=\mathrm{Z} \, \mathrm {B}\, \pmb{\theta}(t) + \pmb{\varepsilon}(t), 
    \end{equation}\bigskip

\noindent where $\mathrm {Z}$ is the $N \times (1+K^1+...+K^p)$ matrix given by
\begin{equation}\label{eq:Z}
    \mathrm {Z}= \int \mathcal {X}(s) \mathrm {H}(s)^T \,ds.
\end{equation}

\section{Estimation of the model}\label{sect:estim}
 In this section we derive an estimator of the unknown functional coefficients of model \eqref{model1mat2}, 
 and we obtain the expression of its variance. 
This is equivalent to estimating the unknown super matrix $\mathrm{B}$ of coefficients in \eqref{model1mat3}; 
to this aim we minimize the Integrated Sum of Square of Errors (ISSE), and then we define the matrix estimator
\begin{equation}\label{eq:defB}
    \hat{\mathrm{B}}=\arg \min_{\mathrm{B}} \mbox{ISSE},
\end{equation}
where
\begin{eqnarray}
      \mbox{ISSE} &=& \int \| \mathbf{y}(t) - \mathrm{Z} \, \mathrm {B}\, \pmb{\theta}(t) \|^2 dt \nonumber \\
      &=& \int \tr(\mathbf{y}(t) - \mathrm{Z} \, \mathrm {B}\, \pmb{\theta}(t))^T (\mathbf{y}(t) - \mathrm{Z} \, \mathrm {B}\, \pmb{\theta}(t)) dt \nonumber \\
      &=& \tr \int (\mathbf{y}(t)^T\mathbf{y}(t) + \pmb{\theta}(t)^T \mathrm{B}^T\, \mathrm{Z}^T\, \mathrm{Z}\, \mathrm{B}\, \pmb{\theta}(t) - 2 \pmb{\theta}(t)^T \mathrm{B}^T\, \mathrm{Z}^T\, \mathbf{y}(t)) dt \nonumber \\
      &=& \tr \int \mathbf{y}(t)^T\mathbf{y}(t) dt + \tr \mathrm{B}^T\, \mathrm{Z}^T\, \mathrm{Z}\, \mathrm{B}\, \mathrm{J}_{{\theta}{\theta}} - 2 \tr \mathrm{B}^T\, \mathrm{Z}^T\, \int \mathbf{y}(t)\, \pmb{\theta}(t)^T dt \label{eq:trISSE};
\end{eqnarray}
 the last inequality follows from the cyclic property of the trace and by defining
\begin{equation*}
    \mathrm{J}_{{\theta}{\theta}}=\int \pmb{\theta}(t)\, \pmb{\theta}(t)^T\, dt.
\end{equation*}

\noindent To obtain the estimator $\hat{\mathrm{B}}$, we calculate the matrix derivatives of \eqref{eq:trISSE} and set them equal to zero; according to the properties of matrix calculus we have:
\begin{eqnarray*}
    \dfrac{\partial}{\partial \mathrm{B}} \tr \mathrm{B}^T\, \mathrm{Z}^T\, \mathrm{Z}\, \mathrm{B}\, \mathrm{J}_{{\theta}{\theta}} &-& 2 \dfrac{\partial}{\partial \mathrm{B}} \tr \mathrm{B}^T\, \mathrm{Z}^T\, \int \mathbf{y}(t)\, \pmb{\theta}(t)^T dt\\
    &=& 2\, \mathrm{Z}^T\, \mathrm{Z}\, \mathrm{B}\, \mathrm{J}_{{\theta}{\theta}} - 2\,  \mathrm{Z}^T\, \int \mathbf{y}(t)\, \pmb{\theta}(t)^T dt = 0,
\end{eqnarray*}
and hence we obtain the normal equations:
\begin{equation*}\label{eq:BISSE}
  \mathrm{Z}^T \mathrm{Z}\, {\mathrm{B}}\,  \mathrm{J}_{{\theta}{\theta}} = \mathrm{Z}^T  \int \mathbf{y}(t) \, \pmb{\theta}(t)^T\, dt.
\end{equation*}

\noindent Since the basis $\{\theta_l(t), l\ge 1\}$ is orthonormal, then $\mathrm{J}_{{\theta}{\theta}}= \mathrm{I}_L$;
if $\mathrm{Z}$ in \eqref{eq:Z} has maximum rank $(1+K^1+...+K^p)$, that is, $\mathrm{Z}^T \mathrm{Z}$ is invertible,
we have the following expression of the estimator \eqref{eq:defB}:
\begin{equation}\label{eq:estimator}
   \hat{\mathrm{B}} =   (\mathrm{Z}^T \mathrm{Z})^{-1}\mathrm{Z}^T  \int \mathbf{y}(t) \, \pmb{\theta}(t)^T\, dt.
\end{equation}


 The following result provides the variance of the vectorization of the matrix estimator \eqref{eq:estimator}. Given a $p\times q$ matrix $\mathrm{M}$ , $\vect(\mathrm{M})$ is the $pq$-vector obtained by stacking the columns of $\mathrm{M}$; the symbol $\otimes$ denotes the Kronecker product on two matrices.
\bigskip

    \begin{proposition}\label{prop:var}
   The estimator \eqref{eq:estimator} is an unbiased estimator of the matrix $\mathrm{B}$ of coefficents of $\pmb{\beta}^*(s,t)$, and its variance is
    \begin{equation*}
        Var(\vect{\mathrm{\hat{B}}})= \mathrm{\Sigma} \otimes (\mathrm{Z}^T \mathrm{Z})^{-1}.
    \end{equation*}
    where $\mathrm{\Sigma}$ is a $L\times L$ matrix whose elements are
    $$\mathrm{\Sigma}_{il}=Cov\left(e_{ni}, e_{nl}\right),$$
    not depending on $n$, with $e_{ni}$ and $e_{nl}$ being the $i$-th and the $l$-th coefficients, respectively, of the expansion of 
    $\varepsilon_n(t)$ in the basis $\{\theta_l(t), l=1, ..., L\}$.
    \end{proposition}
    \begin{proof} The unbiasedness follows from the interchangeability between expectation and integration, and because $\mathrm{J}_{{\theta}{\theta}}= \mathrm{I}_L$, from the orthonormality of $\{\theta_l(t), l=1, ..., L\}$.       

       By considering the representation $\mathbf{y}(t)= \mathrm{Y} \pmb{\theta}(t)$ of the response vector, where $\mathrm{Y}$ is a $N \times L$ matrix of coefficients,  
       we have that 
       \begin{equation*}
           \int \mathbf{y}(t) \, \pmb{\theta}(t)^T\, dt = \mathrm{Y}\, \mathrm{J}_{\theta\theta}=\mathrm{Y},
       \end{equation*}
       and thus the estimator \eqref{eq:estimator} can be expressed as $ \hat{\mathrm{B}} =   (\mathrm{Z}^T \mathrm{Z})^{-1}\mathrm{Z}^T \mathrm{Y}$.
       Note that the random coefficients in each $n$-th row of $\mathrm{Y}$, for $n=1, ..., N$, are dependent between them, since they expand the same  functional response $y_n(t)$; while each row is independent on the others since each row expands a different response.
        In order to obtain an expression of the variance of estimator \eqref{eq:estimator}, it is then convenient to concatenate the rows of $\mathrm{Y}$ and, to get it, to vectorize its transpose ${\hat{\mathrm{B}}}^T$: 
       \begin{equation*}
           {\hat{\mathrm{B}}}^T =  \mathrm{Y}^T \mathrm{Z}\,(\mathrm{Z}^T \mathrm{Z})^{-1} .
       \end{equation*}

  \noindent If $\mathrm{\Phi}$ and $\mathrm{\Psi}$ are a $p\times q$ and an $r\times s$ matrices, respectively, it is possible to show that
       \begin{equation*}\label{eq:vecp}
           \vect(\mathrm{\Phi}\mathrm{\Psi})=(\mathrm{\Psi}^T\otimes \mathrm{I}_p)\vect(\mathrm{\Phi}).
       \end{equation*} 
       Hence, 
  \begin{equation*}
      \vect {{\hat{B}}}^T= [( (\mathrm{Z}^T \mathrm{Z})^{-1}\,\mathrm{Z}^T)\otimes \mathrm{I}_L ] \vect(\mathrm{Y}^T)
  \end{equation*}
  and 
    \begin{equation*}
         Var(\vect{\mathrm{\hat{B}}}^T)=\mathrm{F} \, \mathrm{\Delta} \, \mathrm{F^T},
    \end{equation*}
    where 
\begin{equation*}
  \mathrm{F}=[( (\mathrm{Z}^T \mathrm{Z})^{-1}\, \mathrm{Z}^T)\otimes \mathrm{I}_L ] \quad \mbox{and} \quad \mathrm{\Delta}=Var(\vect \mathrm{Y}^T),  
\end{equation*}

    \noindent which are, respectively, a $(1+\sum_{i=1}^p K^i)\, L \times NL$ and a $NL \times NL$ matrices. 
    To obtain the matrix $\Sigma$, observe that, since $y_1(t), \dots, y_N(t)$ are independent and identically distributed stochastic processes, then the columns of $\mathrm{Y}^T$, that is, $\int y_1(t)\,\pmb{\theta}(t)^T\, dt, ..., \int y_N(t)\,\pmb{\theta}(t)^T\, dt,$ are i.i.d. random vectors. Then,
    $$\Delta= I_N \otimes \Sigma,$$
    where
    \begin{equation}\label{eq:Delta}
       \Sigma_{il}=Cov\left(\int y_n(t)\theta_i(t)\, dt, \int y_n(t)\theta_l(t)\, dt\right) 
    \end{equation}
    which does not depend on $n$.
As a consequence of model \eqref{model1}, the second term in  \eqref{eq:Delta} is equal to
    \begin{equation*}
        Cov\left(\int \varepsilon_n(t)\theta_i(t)\, dt, \int \varepsilon_n(t)\theta_l(t)\, dt\right);
    \end{equation*}
moreover, for any $n$ and $i$,
\begin{equation*}
    \int \varepsilon_n(t)\theta_i(t)\, dt =  \int \sum_{l=1}^L e_{nl} \theta_l(t)\theta_i(t)\, dt =  \sum_{l=1}^L e_{nl}\int \theta_l(t)\theta_i(t)\, dt = e_{ni},
\end{equation*}
where the last equation follows from the orthonormality of $\{\theta_l, l=1, ..., L\}$. 
We finally obtain
\begin{eqnarray}\label{eq;varT}
      Var(\vect{\mathrm{\hat{B}^T}})&=&  [( (\mathrm{Z}^T \mathrm{Z})^{-1}\mathrm{Z}^T)\otimes \mathrm{I}_L ]\, (\mathrm{I}_N \otimes \mathrm{\Sigma})\,\nonumber
    [( \mathrm{Z}\,(\mathrm{Z}^T \mathrm{Z})^{-1})\otimes \mathrm{I}_L ] \label{eq:var} \\
    &=& (( (\mathrm{Z}^T \mathrm{Z})^{-1}\mathrm{Z}^T) \otimes \mathrm{\Sigma})\,
    [( \mathrm{Z}\,(\mathrm{Z}^T \mathrm{Z})^{-1})\otimes \mathrm{I}_L ] \nonumber \\
    &=& (\mathrm{Z}^T \mathrm{Z})^{-1}\otimes \mathrm{\Sigma}. 
    \end{eqnarray}
The two last equalities are due the following property of the Kronecker product:
    if $\mathrm{A}$, $\Phi$, $\Psi$ and $\Omega$ are matrices of sizes that can form the matrices products $\mathrm{A}\, \Psi$ and $\Phi\, \Omega$, then
\begin{equation*}\label{eq:prK}
    (\mathrm{A} \otimes \Phi)(\Psi \otimes \Omega)=(\mathrm{A}\, \Psi)\otimes (\Phi\, \Omega).
\end{equation*}
    
Finally, to obtain $Var(\vect{\mathrm{\hat{B}}})$ from the \eqref{eq;varT}, 
let us denote by $K^{(p,q)}$ the commutation matrix of dimension $pq \times pq$ 
such that $\vect(M^T)=K^{(p,q)}\vect(M)$ for any $p \times q$ matrix $M$ (see \cite[Chapt. 18]{Matrix2019}). Then 
\begin{equation*}
    \vect({\mathrm{\hat{B}}})=K^{(L,1+\sum_{i=1}^p K^i)}\vect({\mathrm{\hat{B}^T}}),
\end{equation*}
and hence
\begin{eqnarray*}
        Var(\vect{\mathrm{\hat{B}}})&=&{K^{(L,1+\sum_{i=1}^p K^i)}((\mathrm{Z}^T \mathrm{Z})^{-1}\otimes \mathrm{\Sigma})K^{(1+\sum_{i=1}^p K^i,L)}}\label{eq:comm}\\
        &=& \mathrm{\Sigma} \otimes  (\mathrm{Z}^T \mathrm{Z})^{-1}, \label{eq:commK}
    \end{eqnarray*}
    where the first equality is justified by the property ${K^{(p,q)}}^T=K^{(q,p)}$,
while the final equality follows from the property of the commutation matrices with Kronecker product (see again \cite[Chapt. 18]{Matrix2019}).    
\end{proof}




\section{Optimal experimental designs}\label{sec:OptDes}

An experimental design for the model described in Section \ref{sec:model}, is defined as a set of experimental conditions $x_{n1}(s), \dots, x_{np}(s)$ in the interval $[0,{\cal T}]$ chosen by the experimenter for each run $n=1, \dots, N$, according to a specific experimental objective; 
 an experimental design is then identified by the design matrix $\mathrm{X}(s)$ in \eqref{eq:Dmatrix}. 
Once a basis of functions is defined, we can expand these experimental conditions:
\begin{equation}\label{eq:expxni}
    x_{ni}(s)={\pmb{\gamma}_{n}^i}^T\, \pmb{\eta}^i(s),
\end{equation}
 for any $n=1, ..., N$ and $i=1, ..., p$, where $\pmb{\gamma}_{n}^i$ is a $K^i$-dimensional vector of coefficients. 
 Recalling that the model matrix \eqref{eq:Xmatrix} is obtained by adding a column of one's to the matrix of the experimental conditions,  we have that it can be rewritten as
 \begin{equation*}
     \mathcal{X}(s)= \Gamma \mathrm {H}(s),
 \end{equation*}

 \noindent where $\Gamma$ is the $N \times (1+ K^1+...+K^p)$ block matrix given by
\begin{equation}\label{eq:Gamma}
    \Gamma=(\mathbf{1}_N \,|\,\Gamma^1 \, | \, \hdots \,| \, \Gamma^p),
\end{equation}
with \begin{equation*}
    \Gamma^i=\begin{bmatrix}
    {\pmb{\gamma}_{n}^i}^T\\
    \vdots \\
    \\
{\pmb{\gamma}_{N}^i}^T
    \end{bmatrix}
\end{equation*}
Hence,  once a basis of function is chosen, an experimental design is identified by the matrix $\Gamma$ in \eqref{eq:Gamma}.
\bigskip

We can then compute the matrix $Z$ defined in \eqref{eq:Z}, obtaining

\begin{equation}\label{eq:ZJHH}
   \mathrm{Z}=  \Gamma\, \int \mathrm {H}(s)\, \mathrm{ H}(s)^T \,ds 
            = \Gamma\, \mathrm{J}_{HH}
\end{equation}
where
\begin{equation}\label{eq:JHH}
    \mathrm{J}_{HH}= \mbox{diag}(1, \mathrm{J}_{{\eta}{\eta}^1}, \dots, \mathrm{J}_{{\eta}{\eta}^p})
\end{equation}
and, for any $i=1, \dots, p$,
\begin{equation*}
    \mathrm{J}_{{\eta}{\eta}^i}=\int \pmb{\eta}^i(s)\, \pmb{\eta}^i(s)^T\, ds.
\end{equation*}
\bigskip

The following definition extends the classical definition of the optimal design literature (see for instance \cite{Atk2007}) to the functional framework considered in this paper. Since we have proved that the estimator considered is unbiased, then minimizing a suitable functional of its variance allows the most precise estimation of the unknown functional parameters of model \eqref{model1mat}.

\begin{definition}\label{def:optd}
    An optimal design  is an experimental design $X^*(s)$ which minimizes a proper function $\Phi$ of the variance of the estimator ${\mathrm{\hat{B}}}$ in \eqref{eq:estimator}:
\begin{equation*}\label{eq:opt}
    X^*(s)=\arg \min_{X(s)} \Phi (Var(\vect {\mathrm{\hat{B}}})).
\end{equation*}
\end{definition} 


When the basis to expand the factors and their correspondent coefficients are the same, Definition \ref{def:optd} means that an optimal design $X^*(s)$ is identified by a matrix $\Gamma^*$ such that
\begin{equation*}
    \Gamma^*=\arg \min_{\Gamma} \Phi (\mathrm{\Sigma} \otimes (\mathrm{Z}^T \mathrm{Z})^{-1})=
    \arg \min_{\Gamma} \Phi (\mathrm{\Sigma} \otimes ({\mathrm{J}_{HH}}^T\Gamma^T\, \Gamma\, \mathrm{J}_{HH})^{-1}).
\end{equation*}

\subsection{Use of different bases to expand parameters and factors}\label{sec:diff_bas}
In the experimental context, sometimes it could be useful to expand each factor with a different basis $\{c^i_k(s), 1 \le k\le K_x^i\}$ than the basis $\{\eta^i_k(s), 1 \le k\le K^i\}$ used for 
the unknown coefficients $\beta_i^*(s,t)$, $i=1, \dots, p$.  For instance, a step functions basis could be the favourite choice by the practitioner, while a different basis could be chosen for representing the functional coefficient. Our methodology is flexible to this option, and it is possible also to consider different dimensions $K_x^i\neq K^i$. In this situation, we write
\begin{equation*}
    x_{ni}(s)={\pmb{\gamma}_{n}^i}^T\, \pmb{c}^i(s),
\end{equation*}
 for any $n=1, ..., N$ and $i=1, ..., p$, where ${\pmb{\gamma}_{n}^i}$ is a $K_x^i$-dimensional vector of coefficients. Then,
 \begin{equation*}
     \mathcal {X}(s)= \Gamma\, \mathrm {C}(s)\, ds,
 \end{equation*}
where
\begin{equation*}
    \mathrm{C}(s)=\mbox{diag}(1, \pmb{c}^1(s), \hdots,\, \pmb{c}^p(s)),
\end{equation*}

 \noindent and again $\Gamma$ is a $N \times (1+ K_x^1+...+K_x^p)$ block matrix given by \eqref{eq:Gamma}.
Hence,

\begin{equation}\label{eq:ZJCH}
   \mathrm{Z}=  \Gamma\, \int \mathrm {C}(s)\, \mathrm{ H}(s)^T \,ds 
            = \Gamma\, \mathrm{J}_{CH}
\end{equation}
where
\begin{equation}\label{eq:JCH}
    \mathrm{J}_{CH}= \mbox{diag}(1, \mathrm{J}_{{c}{\eta}^1}, \dots, \mathrm{J}_{{c}{\eta}^p})
\end{equation}
and, for any $i=1, \dots, p$,
\begin{equation*}
    \mathrm{J}_{{c}{\eta}^i}=\int \pmb{c}^i(s)\, \pmb{\eta}^i(s)^T\, ds.
\end{equation*}

In this case, Definition \ref{def:optd} means that an optimal design $X^*(s)$ is identified by a matrix $\Gamma^*$ such that
\begin{equation*}
    \Gamma^*=\arg \min_{\Gamma} \Phi (\mathrm{\Sigma} \otimes (\mathrm{Z}^T \mathrm{Z})^{-1})=
    \arg \min_{\Gamma} \Phi (\mathrm{\Sigma} \otimes ({\mathrm{J}_{CH}}^T\Gamma^T\, \Gamma\, \mathrm{J}_{CH})^{-1}).
\end{equation*}

\subsection{A- and D-optimality}

In this section, we consider the A- and D-optimality criteria (see again \cite{Atk2007}) for the precise estimation of the functional coefficient. We extend these criteria to dynamic experimental conditions:
a functional A-optimal design $X^*_A(s)$ and a functional D-optimal design $X^*_D(s)$
minimize  $\Phi (Var(\vect {\mathrm{\hat{B}}}))$, according to Definition \ref{def:optd},
with $\Phi=\tr$, 
and  $\Phi=\det$, respectively. 

In the following propositions we derive the expression of these criteria, and in particular we prove that they do not depend on the unknown covariance matrix $\Sigma$ of the error process. These results allow the computation of A-optimal and D-optimal design, as it will be exemplified in Section \ref{sec:ex}.


\bigskip

\begin{proposition} A functional A-optimal design $X^*_A(s)$ is obtained by minimizing $\tr(\mathrm{Z}^T \mathrm{Z})^{-1}$,
    where $Z$ is the matrix defined in \eqref{eq:Z}. 
\end{proposition}
\begin{proof}
    From Proposition \ref{prop:var} and from the property of Kronecker product that
    \begin{equation*}
        \tr (A \otimes B) = (\tr A)( \tr B),
    \end{equation*}
    we have that
\begin{equation}\label{eq:tr}
         \tr Var(\vect{\mathrm{\hat{B}}})=\tr \Sigma\, \tr(\mathrm{Z}^T \mathrm{Z})^{-1};
    \end{equation}
note that $\Sigma$ is the covariance matrix of the error process and it is not affected by the design;  then, if a functional design  
minimizes $\tr(\mathrm{Z}^T \mathrm{Z})^{-1}$, it minimizes also the \eqref{eq:tr}, and hence it is A-optimal.
\end{proof}

\begin{proposition} A functional D-optimal design $X^*_D(s)$ is obtained by maximizing ${ \det \mathrm{Z}^T \mathrm{Z}}$,
    where $Z$ is defined in \eqref{eq:Z}. 
\end{proposition}
\begin{proof}
    Again, from Proposition \ref{prop:var} and from the property of Kronecker product that if A and B are an $n\times n$ and a $m\times m$ matrices, respectively, then
    \begin{equation*}
        \det(A \otimes B)=(\det A)^m (\det B)^n,
    \end{equation*}
    we have that
\begin{equation}\label{eq:det}
    \det Var(\vect{\mathrm{\hat{B}}})
    =\dfrac{(\det \mathrm{\Sigma})^{1+\sum_{i=1}^p K^i}}{(\det \, \mathrm{Z}^T \mathrm{Z})^{L}};
\end{equation}
then, if a functional design  
maximizes $\det \, \mathrm{Z}^T \mathrm{Z}$,  it minimizes the \eqref{eq:det}, and hence it is D-optimal.
\end{proof}

Note that, for the computation of the optimal experimental designs through an optimization algorithm, the matrix $Z$ can be obtained from \eqref{eq:ZJHH}, whenever the bases for expanding the first variable of the functional coefficient and the factors are the same, or by \eqref{eq:JCH}, whenever the bases are different. The optimal designs will depend on the original choice of the bases.

\bigskip

\section{Results: optimal designs}\label{sec:ex}
This section is devoted to offer some examples of functional optimal experimental designs for precise estimation of the functional coefficient of model \eqref{model1mat2}, based on the theory developed in the previous sections. The examples are organized as follows: first, we consider the same basis of functions to expand the functional coefficient and the runs of one functional predictor (Subsection \ref{ex:scenario1}); then, we extend to the case of choosing two different bases (Subsection \ref{ex:scenario2}); finally, we present an example with more than one predictor (Subsection \ref{ex:scenario3}).  

In all of the examples various B-splines \cite{smith1980practical} are used as a choice for expanding the basis for both the functional factors and parameters. A B-spline is defined by its degree $D$ and its knot vector. If we assume that the knot vector is constructed by equidistant points in the interval $[0,1]$, then we can define it in terms of the number of breakpoints $\mathcal{K}$ (in our notation including the two bound points 0 and 1). For example the knots vector with $\mathcal{K}=5$ should be $\{0, 0.25, 0.5, 0.75, 1\}$. The length for a basis of degree $D$, using     $\mathcal{K}$ knots in the knot vector, is $D + \mathcal{K} - 1$. 

In the examples with one functional predictor we assume that $N=12$ runs are available and we investigate how increasing the number of breakpoints for expanding the functional factor $\mathbf{x}(s)$ affects the optimality. Additionally, we increase the number of break points in the expansion of $\beta(s,t)$ with respect to $s$, to investigate how the design output chances for a constant number of breakpoints in $X(s)$. For a design to be estimable we need the total number of parameters (plus one to include a constant term according to \eqref{eq:Xmatrix}) to be less than the number of experimental runs. So if we expand $\beta(s,t)$ with a B-spline of degree $D_\beta$ and $\mathcal{K}_\beta$ breakpoints we need $N > D_\beta + \mathcal{K}_\beta$. Additionally, if we use a B-spline of degree $D_X$ and $\mathcal{K}_X$ knots for expanding the functional factor $X(s)$ we need to enforce $D_X + \mathcal{K}_X \geq D_\beta + \mathcal{K}_\beta$, to ensure identifiability of \eqref{model1mat2}.

Additionally, we explore scenarios involving multiple functional predictors to demonstrate how the interplay between different factors and their respective basis expansions influences the optimal design. Specifically, Subsection \ref{ex:scenario3} presents an example with more than one predictor, illustrating the complexities and considerations that arise when extending the design framework to accommodate multiple functional inputs.

The algorithm used is a version of the coordinate-exchange \cite{meyer1995coordinate} using $1000$ random starts implemented in Python 3.10  \cite{van1995python}. The code is available by the authors upon request.
The examples are referred to A-optimality, similarly D-optimum designs could be obtained.

\subsection{Same basis of functions to expand the dynamic factor and the parameter}\label{ex:scenario1}
In this scenario, we consider one functional factor and we expand both $X(s)$ and the functional parameter $\beta(s,t)$ using zero-degree B-spline basis functions ($D_X = 0$ and $D_\beta = 0$). Zero-degree B-splines are defined over a set of equidistant breakpoints (or knots) within the domain of the function. Between each pair of consecutive breakpoints, the basis function is constant.

In our analysis, we investigate how varying the number of breakpoints in the B-spline basis for $X(s)$ affects the A-optimality criterion value while keeping the basis and number of breakpoints of $\beta(s,t)$, with respect to the $s$ variable, fixed. More specifically, we assume $\mathcal{K}_X \in \{3, 5, 9, 15, 19, 29\}$ and $\mathcal{K}_\beta \in \{3, 5 ,7\}$ and we vary the number of breakpoints $\mathcal{K}_X$ for each fixed $\mathcal{K}_\beta$.

Table \ref{tab:scen-1} summarizes the A-optimality values and the relative efficiencies for different combinations of $\mathcal{K}_X$ and $\mathcal{K}_\beta$. The computed efficiency is  relative to the best A-optimality value achieved in each case. 

\begin{table}[h]
    \centering
    \begin{tabular}{c S[table-format=2.2] S[table-format=1.2] S[table-format=2.2] S[table-format=1.2] S[table-format=2.2] S[table-format=1.2]}
        \toprule
        \textit{Breaks} &
        \multicolumn{2}{c}{\textit{3}} &
        \multicolumn{2}{c}{\textit{5}} &
        \multicolumn{2}{c}{\textit{7}} \\
        \cmidrule(lr){2-3} \cmidrule(lr){4-5} \cmidrule(lr){6-7}
        & {A-opt} & \textbf{Efficiency} &
          {A-opt} & \textbf{Efficiency} &
          {A-opt} & \textbf{Efficiency} \\
        \midrule
        \textit{3}   & 0.75 & 1.00 & {-}  & {-}  & {-}     & {-}    \\
        \textit{5}   & 0.75 & 1.00 & 5.42 & 1.00 & {-}     & {-}    \\
        \textit{9}   & 0.75 & 1.00 & 5.42 & 1.00 & 27.25   & 0.69   \\
        \textit{15}  & 0.75 & 1.00 & 6.33 & 0.86 & 23.46   & 0.80   \\
        \textit{19}  & 0.75 & 1.00 & 6.10 & 0.89 & 18.70   & 1.00   \\
        \textit{29}  & 0.75 & 1.00 & 5.42 & 1.00 & 21.04   & 0.89   \\
        \bottomrule
    \end{tabular}
    \caption{A-optimality values and relative efficiencies for zero-degree B-splines expansion of $X(s)$ and $\beta(s,t)$ with varying numbers of breakpoints in $X(s)$.}
    \label{tab:scen-1}
\end{table}

From Table \ref{tab:scen-1}, we observe that when $\mathcal{K}_\beta = 3$, the efficiency remains constant at 100\% no matter the number of breakpoints for $X(s)$. This indicates that increasing the number of breakpoints of the predictor's basis functions does not improve the estimation precision of $\beta(s,t)$ when the coefficient function is represented with only two ($D_\beta=0$, $\mathcal{K}_\beta=3$) basis functions. 

\begin{figure}[htbp]
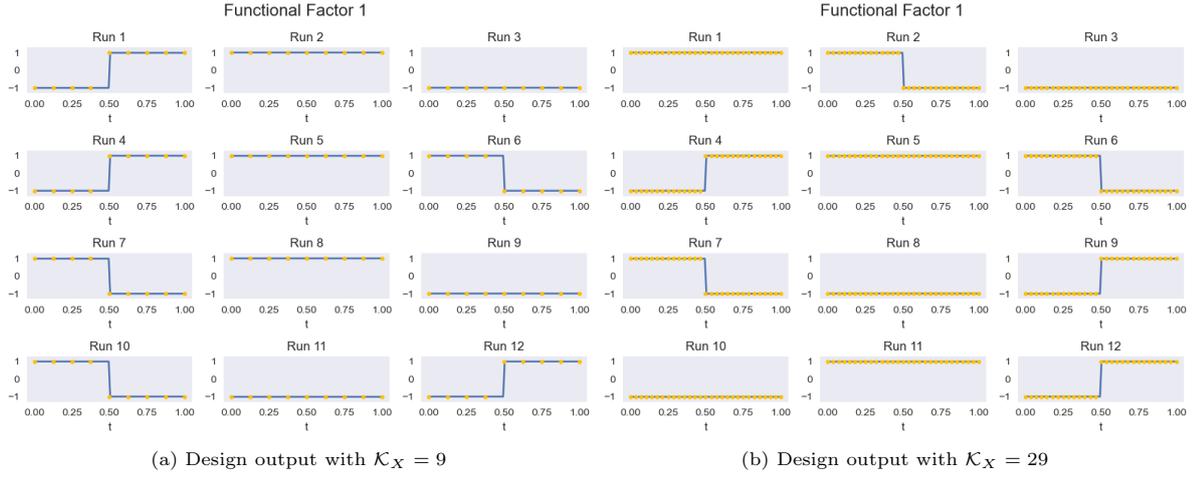

    \makebox[\textwidth][c]{%
        \begin{minipage}{1.2\textwidth}
            \centering
            \subfloat[Design output with $\mathcal{K}_X = 9$]{%
                \includegraphics[width=0.5\textwidth]{assets/main/scenario_1/beta_3/1000_100_12_bpslines_0_9_bsplines_0_3.pdf}%
                \label{fig:scen-1.1-a}}
            \hfill
            \subfloat[Design output with $\mathcal{K}_X = 29$]{%
                \includegraphics[width=0.5\textwidth]{assets/main/scenario_1/beta_3/1000_100_12_bpslines_0_29_bsplines_0_3.pdf}%
                \label{fig:scen-1.1-b}}
        \end{minipage}
    }
    \caption{Comparison of design output for different breakpoints when $\mathcal{K}_\beta = 3$.}
    \label{fig:scen-1.1}
\end{figure}

This observation is further illustrated in Figure \ref{fig:scen-1.1}, where we compare the optimal design outputs for $\mathcal{K}_X = 15$ and $\mathcal{K}_X = 29$ with $\mathcal{K}_\beta = 3$. Despite the increase in the number of breakpoints for $X(s)$, the overall shape of the optimal design remains unchanged. Essentially, the simplicity of the coefficient function limits the complexity that can be captured in the design, and thus, even with more breakpoints, the optimal design does not change. 

In contrast, when $\mathcal{K}_\beta = 7$, allowing for a more flexible representation of the factor by increasing the number of breakpoints, improves the efficiency of the design, albeit with diminishing returns. A more flexible basis for $X(s)$ allows the design to capture the increased complexity of the more intricate coefficient function $\beta(s,t)$, enhancing the estimation precision.

\begin{figure}[htbp]
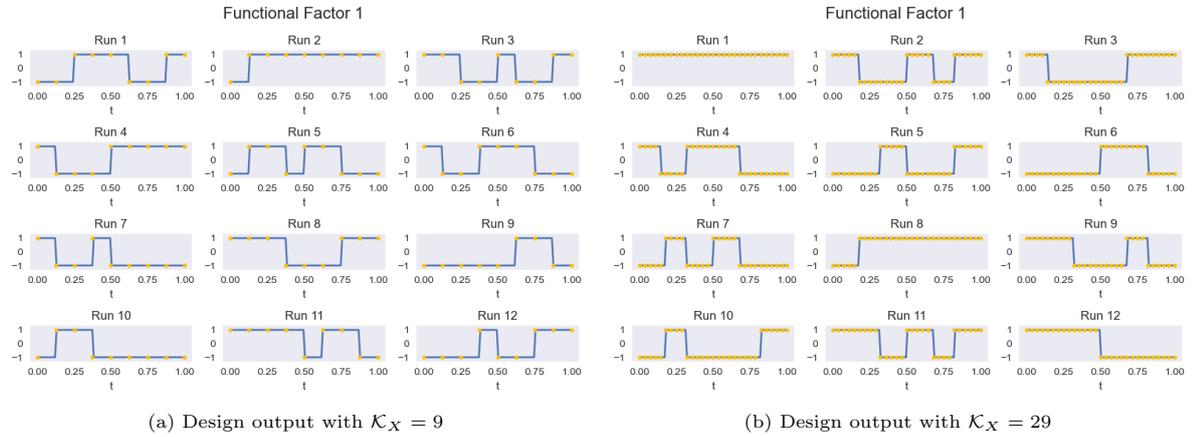

\makebox[\textwidth][c]{%
        \begin{minipage}{1.2\textwidth}
    \centering
    \subfloat[Design output with $\mathcal{K}_X = 9$]{%
        \includegraphics[width=0.495\linewidth]{assets/main/scenario_1/beta_7/1000_100_12_bpslines_0_9_bsplines_0_7.pdf}%
        \label{fig:scen-1.2-a}}
    \hfill
    \subfloat[Design output with $\mathcal{K}_X = 29$]{%
        \includegraphics[width=0.495\linewidth]{assets/main/scenario_1/beta_7/1000_100_12_bpslines_0_29_bsplines_0_7.pdf}%
        \label{fig:scen-1.2-b}}
        \end{minipage}
    }
    \caption{Comparison of design output for different breakpoints when $\mathcal{K}_\beta = 7$.}
    \label{fig:scen-1.2}
\end{figure}

This can be seen in the design output in figure \ref{fig:scen-1.2}. Increasing the number of breakpoints for $X(s)$ from 9 to 29 changes the shapes. Upon closer inspection of figure \ref{fig:scen-1.2} we see that there is no run in the left panel that matches in shape any from the right panel. This reinforces our belief that a more complicated expansion for the functional factor is better at capturing a more complicated expansion of the functional parameter.

An interesting observation arises from Table \ref{tab:scen-1} concerning the case with 19 breakpoints for $X(s)$ and 7 breakpoints for $\beta(s,t)$. Notably, this specific configuration yields a better A-optimality criterion value than the case with 29 breakpoints for $X(s)$. To understand this why this happens, it is essential to examine the knot vectors of $X(s)$ and $\beta(s,t)$ in this scenario.

We deduce that the knot vector of $\beta(s,t)$ is included within the knot vector of $X(s)$ when 19 breakpoints are used. This inclusion means that every knot of $\beta(s,t)$ aligns with a knot of $X(s)$, facilitating a more precise estimation of the coefficient function.
In contrast, this alignment does not occur for $X(s)$ with 9 or 15 breakpoints, where the step sizes do not result in such inclusion. 
This observation suggests that carefully aligning the knot vectors of $X(s)$ and $\beta(s,t)$ can enhance the estimation precision, and the knot vector of the predictor should be thoughtfully selected based on that of the coefficient function.

A similar rationale applies to the case with 5 breakpoints for $\beta(s,t)$ and 15 or 19 breakpoints for $X(s)$. 
In these scenarios, the knot vector of the coefficient function is not included within that of the predictor. As a result, the design cannot achieve 100\% efficiency compared to cases where such inclusion exists.


These findings suggest the importance of matching not only the number of knots of the predictor's basis functions with the complexity of the coefficient function but also ensuring the alignment of their knot vectors. When the coefficient function is more intricate, increasing the number of knots of the predictor's basis functions is beneficial. However, optimal efficiency is further achieved when the knot vectors are carefully constructed so that the knots of the coefficient function are included within those of the predictor. This alignment allows the predictor to capture the essential features of the coefficient function more effectively, leading to improved estimation precision.

In the next section, we will explore scenarios where different families of basis functions are employed for the functional predictor and the parameter. This investigation will further elucidate how the choice of basis functions and the alignment of knot vectors impact the efficiency and effectiveness of the experimental design.

\subsection{Different bases of functions}\label{ex:scenario2}
In this scenario, we explore the optimal experimental design when the functional predictor $X(s)$ and the functional coefficient $\beta(s,t)$ are expanded using different bases functions. Specifically, we consider cases where the predictor and the coefficient function are represented using B-splines of different degrees. This investigation aims to understand how the choice of basis functions for the predictor and the coefficient affects the efficiency of the design.

We focus on two sub-scenarios:

\begin{enumerate}
    \item $D_X=1$ and $D_\beta = 2$: This setup allows us to assess the impact of having a more complex coefficient function relative to the input factor.
    \vspace{0.1cm}
    \item $D_X=1$ and $D_\beta=0$: This contrast helps us understand the effect when the factor's flexibility exceeds that of the coefficient function.
\end{enumerate}

In both sub-scenarios, we vary the number of breakpoints (knots) in the B-spline basis functions for $X(s)$ to observe how increasing the number of knots of the predictor influences the A-optimality criterion and the efficiency of the design.

\subsubsection{First-degree B-splines for $X(s)$ and Second-degree B-splines for $\beta(s,t)$}

In this sub-scenario, we investigate the optimal design when the functional predictor $X(s)$ is expanded using first-degree B-spline basis functions (piecewise linear functions), and the coefficient function $\beta(s,t)$ is expanded using second-degree B-spline basis functions (piecewise quadratic functions). This setup allows us to examine the effect of using a basis for the coefficient function that is allowed to change in more points relative to the predictor.

We consider varying the number of breakpoints in the B-spline basis for $X(s)$ ($\mathcal{K}_X \in \{5,9,15,19\}$), while keeping the number of breakpoints for $\beta(s,t)$ fixed at three different values as per the previous example ($\mathcal{K}_\beta \in \{3,5,7\}$). By analyzing both cases, we can assess how the complexity of the coefficient function impacts the efficiency of the design and the estimation precision when paired with varying predictor flexibility.

Table \ref{tab:scen-2.1} summarizes the A-optimality values and relative efficiencies for these settings. The efficiency is calculated relative to the best A-optimality value achieved in each case, which corresponds to the maximum number of breakpoints for $X(s)$.

\begin{table}[h]
    \centering
    \begin{tabular}{c
                    S[table-format=2.2] S[table-format=1.2]
                    S[table-format=2.2] S[table-format=1.2]
                    S[table-format=2.2] S[table-format=1.2]}
        \toprule
        \textit{Breaks} &
        \multicolumn{2}{c}{\textit{3}} &
        \multicolumn{2}{c}{\textit{5}} &
        \multicolumn{2}{c}{\textit{7}} \\
        \cmidrule(lr){2-3} \cmidrule(lr){4-5} \cmidrule(lr){6-7}
        & {A-opt} & \textbf{Efficiency} &
          {A-opt} & \textbf{Efficiency} &
          {A-opt} & \textbf{Efficiency} \\
        \midrule
        \textit{5}   & 36.82  & 0.60  &  {-}    & {-}  & {-}      & {-}    \\
        \textit{9}   & 23.71  & 0.92  & 112.84  & 0.71 & 352.12   & 0.60   \\
        \textit{15}  & 22.49  & 0.98  & 84.44   & 0.95 & 256.99   & 0.79   \\
        \textit{19}  & 22.19  & 0.99  & 83.93   & 0.96 & 216.76   & 0.94   \\
        \textit{29}  & 21.96  & 1.00  & 80.30   & 1.00 & 207.83   & 1.00   \\
        \bottomrule
    \end{tabular}
    \caption{A-optimality values and relative efficiencies for first-degree B-splines expansion of $X(s)$ and second-degree B-splines of $\beta(s,t)$ with varying numbers of breakpoints in $X(s)$.}
    \label{tab:scen-2.1}
\end{table}

\begin{figure}[htbp]
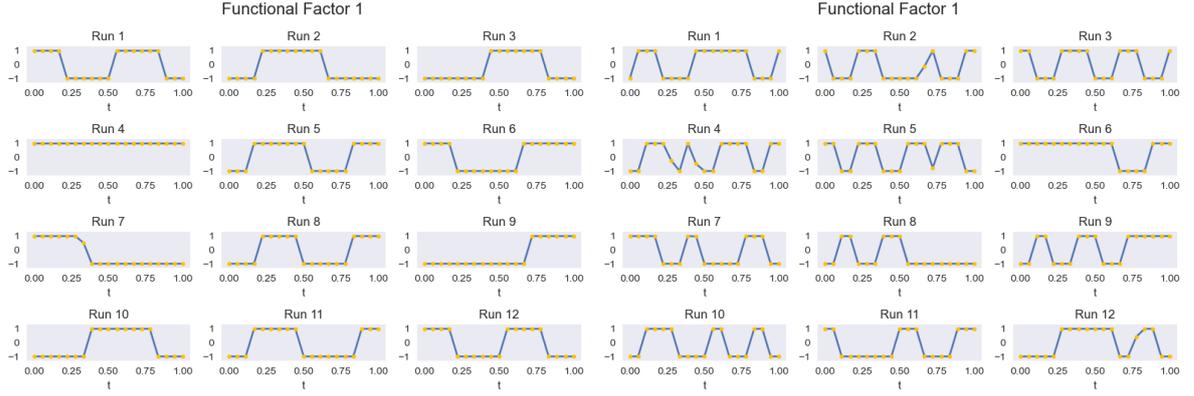

\makebox[\textwidth][c]{%
        \begin{minipage}{1.2\textwidth}
    \centering
    \subfloat[Design output with $\mathcal{K}_\beta = 3$]{\includegraphics[width=0.5\textwidth]{assets/main/scenario_2/beta_3/1000_100_12_bpslines_1_19_bsplines_2_3.pdf}\label{fig:scen-2.1-a}}
    \hfill
    \subfloat[Design output with $\mathcal{K}_\beta = 7$]{\includegraphics[width=0.5\textwidth]{assets/main/scenario_2/beta_7/1000_100_12_bpslines_1_19_bsplines_2_7.pdf}\label{fig:scen-2.1-b}}
    \caption{Comparison of design output for different breakpoints when $\mathcal{K}_X = 19$.}
    \label{fig:scen-2.1}
    \end{minipage}
    }
\end{figure}

In Figure \ref{fig:scen-2.1} we present two designs with a fixed number of breakpoints for $X(s)$ ($\mathcal{K}_X = 19$), for $\mathcal{K}_\beta \in \{3,7\}$. We can observe that by increasing the complexity of the functional parameter significantly impacts the design. The optimal design adapts by utilizing the available knots in $X(s)$ to generate more varied predictor functions for the more complicated $\beta(s,t)$. This is clear when comparing the changes in slope between Figure \ref{fig:scen-2.1-a} and Figure \ref{fig:scen-2.1-b}. Figure \ref{fig:scen-2.1-a} has much fewer changes in slope for each experimental run because the coefficient function is simpler.

This sub-scenario demonstrates that when the predictor's representation is fixed, increasing the complexity of the coefficient function requires the design to make full use of the available knots to achieve optimal estimation. The designs become more intricate, with more changes in slope in the predictor functions, reflecting the need to capture the additional complexity in $\beta(s,t)$.

\subsubsection{First-degree B-splines for $X(s)$ and Zero-degree B-splines for $\beta(s,t)$}\label{ex:scenario2.2}
In this last sub-scenario, we explore the optimal design when the functional predictor $X(s)$ is expanded using first-degree B-spline basis functions, while the coefficient function $\beta(s,t)$ is expanded using zero-degree B-spline basis functions. This setup allows us to examine the effect of using a more flexible basis with respect to the number of breakpoints for the predictor relative to a simpler coefficient function.

We will again vary the number of breakpoints $\mathcal{K}_X$ in the B-spline basis for $X(s)$ while keeping the number of breakpoints for $\beta(s,t)$ fixed. We assess how the predictor's number of knots impacts the efficiency of the design and the estimation precision when the coefficient function is relatively simple.

\begin{table}[h]
    \centering
    \begin{tabular}{c
                    S[table-format=2.2] S[table-format=1.2]
                    S[table-format=2.2] S[table-format=1.2]
                    S[table-format=2.2] S[table-format=1.2]}
        \toprule
        \textit{Breaks} &
        \multicolumn{2}{c}{\textit{3}} &
        \multicolumn{2}{c}{\textit{5}} &
        \multicolumn{2}{c}{\textit{7}} \\
        \cmidrule(lr){2-3} \cmidrule(lr){4-5} \cmidrule(lr){6-7}
         & {A-opt} & \textbf{Efficiency} &
           {A-opt} & \textbf{Efficiency} &
           {A-opt} & \textbf{Efficiency} \\
        \midrule
        \textit{2}   & 1.58  & 0.49  & {-}   & {-} & {-}     & {-}    \\
        \textit{3}   & 1.34  & 0.58  & {-}   & {-} & {-}     & {-}    \\
        \textit{5}   & 0.97  & 0.80  & 15.01 & 0.40 & {-}     & {-}    \\
        \textit{9}   & 0.85  & 0.91  & 8.13  & 0.74 & 32.21   & 0.66   \\
        \textit{15}  & 0.80  & 0.97  & 6.35  & 0.95 & 25.17   & 0.85   \\
        \textit{19}  & 0.79  & 0.98  & 6.11  & 0.99 & 24.69   & 0.86   \\
        \textit{29}  & 0.77  & 1.00  & 6.04  & 1.00 & 21.34   & 1.00   \\
        \bottomrule
    \end{tabular}
    \caption{A-optimality values and relative efficiencies for first-degree B-spline expansion of $X(s)$ and zero-degree B-spline expansion of $\beta(s,t)$ with varying numbers of breakpoints in $X(s)$.}
    \label{tab:scen-2.2}
\end{table}

In Table \ref{tab:scen-2.2} we again observe the same thing as with the previous sections. The efficiency of the design increases for all cases of $\mathcal{K}_\beta$ when we allow $X(s)$ to vary between more breakpoints, with diminishing returns. 

In scenario \ref{ex:scenario1} we assumed the same zero-degree B-spline basis for the expansion of $\beta(s,t)$ as in this scenario. Hence, the A-optimality values presented in Table \ref{tab:scen-2.2} are directly comparable to Table \ref{tab:scen-1} from scenario \ref{ex:scenario1}. The only thing that changed between those two scenarios is the type of basis function chosen for expanding the functional factor. In scenario \ref{ex:scenario1} we used a zero-degree B-splines basis and in this scenario we allow for a more complex first-degree B-splines basis.

In Table \ref{tab:scen-2.2} we can see that even at the maximum number of breakpoints, $\mathcal{K}_X=29$, we never quite reach the same A-optimality values as in Table \ref{tab:scen-1} of scenario \ref{ex:scenario1} with respect to $\mathcal{K}_\beta$. This suggest that the design with the more abrupt changes in slope (scenario \ref{ex:scenario1}), produces a more efficient design that the one with a more gradual change in slope (scenario \ref{ex:scenario2.2}).

\begin{figure}[htbp]
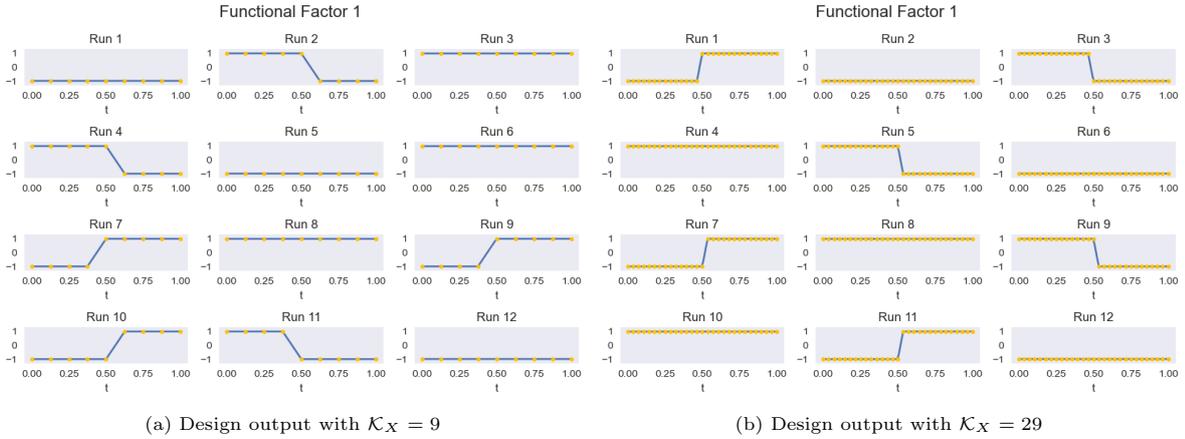

\makebox[\textwidth][c]{%
        \begin{minipage}{1.2\textwidth}
    \centering
    \subfloat[Design output with $\mathcal{K}_X = 9$]{\includegraphics[width=0.5\textwidth]{assets/main/scenario_2/beta_3/1000_100_12_bpslines_1_9_bsplines_0_3.pdf}\label{fig:scen-2.2-a}}
    \hfill
    \subfloat[Design output with $\mathcal{K}_X = 29$]{\includegraphics[width=0.5\textwidth]{assets/main/scenario_2/beta_3/1000_100_12_bpslines_1_29_bsplines_0_3.pdf}\label{fig:scen-2.2-b}}
    \caption{Comparison of design output for different breakpoints when $\mathcal{K}_\beta = 3$.}
    \label{fig:scen-2.2}
    \end{minipage}
    }
\end{figure}

To visualize the impact of the different basis functions we present Figure \ref{fig:scen-2.2}, which shows the optimal designs for $\mathcal{K}_X=9$, $\mathcal{K}_\beta=29$ and $\mathcal{K}_\beta=3$.

By comparing Figure \ref{fig:scen-2.2-a} to Figure \ref{fig:scen-1.1-a} we see that the shapes of each experimental run of the functional factors are the same, given the constraints of each basis. For example either the function experience no change in slope and are pushed to the extreme values of the $[-1,1]$ range that they are allowed to take, or they show one change in slope at the mid-point of the domain (0.5). The zero-degree B-spline basis of Figure \ref{fig:scen-1.1-a} results in more abrupt jumps from the one extreme point to the other, while the smoother first-degree B-spline basis of Figure \ref{fig:scen-2.2-a}, has a more gentle slope going from one extreme to the other. The same can be said for the comparison between Figure \ref{fig:scen-2.2-b} and Figure \ref{fig:scen-1.1-b}. 

This last sub-scenario demonstrates that increasing the predictor's complexity by using higher-degree basis functions does not necessarily improve the estimation precision when the coefficient function is relatively simple. Overall, this analysis underscores that the choice of basis functions for the predictor and coefficient function should be carefully considered in functional experimental design. 

\subsection{More than one predictor}\label{ex:scenario3}
In this final scenario, we extend our investigation to the case of multiple functional predictors. Specifically, we consider a model with two functional predictors $\mathbf{x}_1$ and $\mathbf{x}_2$, each one expanded using different B-spline basis functions. 
This example allows us to explore how the interplay between multiple predictors and their corresponding coefficient functions affects the optimal design and estimation precision.

The predictors and their associated coefficient functions are expanded as summarized in Table~\ref{tab:predictors}.

\begin{table}[h]
\centering
\begin{tabular}{llcc|cc}
\toprule
 & & \multicolumn{2}{c|}{\textit{Component 1}} & \multicolumn{2}{c}{\textit{Component 2}} \\
\cmidrule(lr){3-4} \cmidrule(lr){5-6}
 & & \textbf{Degree} & \textbf{Breakpoints} & \textbf{Degree} & \textbf{Breakpoints} \\
\midrule
\textbf{Factor} & \(\mathbf{x}(s)\) & \(0\) & \(5\) & \(2\) & \(9\) \\
\textbf{Coefficient} & \(\beta(s,t)\) & \(0\) & \(3\) & \(1\) & \(3\) \\
\bottomrule
\end{tabular}
\caption{Degree of the B-spline and number of breakpoints for the two functional factors and coefficient pairs.}
\label{tab:predictors}
\end{table}\vspace{-0.5cm}

In this scenario, we fix the number of breakpoints for both predictors. The A-optimality criterion value obtained for this design is $6.425$.

\begin{figure}[htbp]
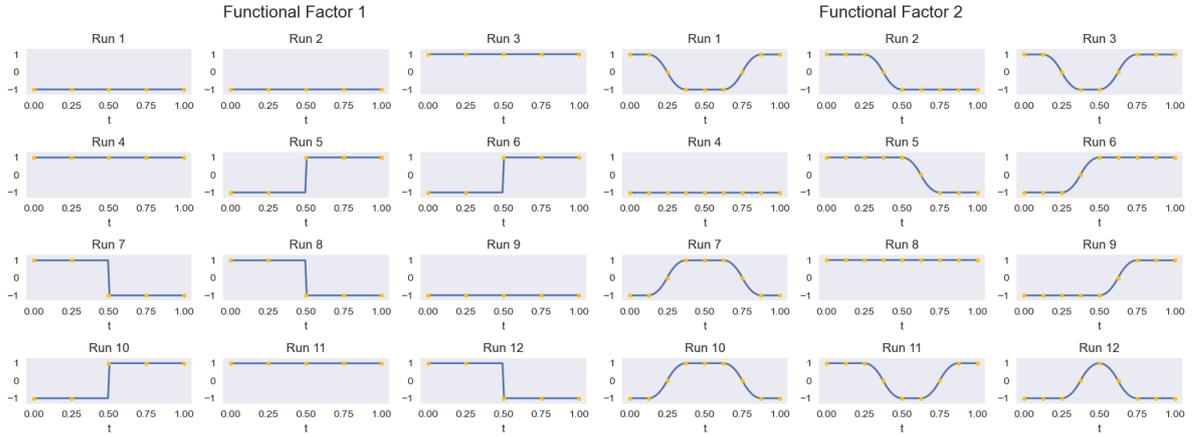

\makebox[\textwidth][c]{%
        \begin{minipage}{1.2\textwidth}
    \centering
    \subfloat[Design output for $X_1(s)$]{\includegraphics[width=0.5\textwidth]{assets/main/scenario_3/1000_100_bsplines_0_5_bsplines_0_3.pdf}\label{fig:scen-3-a}}
    \hfill
    \subfloat[Design output for $X_2(s)$]{\includegraphics[width=0.5\textwidth]{assets/main/scenario_3/1000_100_bsplines_2_9_bsplines_1_3.pdf}\label{fig:scen-3-b}}
    \caption{Design of experiments for two experimental factors.}
    \label{fig:scen-3}
    \end{minipage}
    }
\end{figure}

Figure \ref{fig:scen-3} presents the optimal design for this scenario, displaying the experimental runs for both functional predictors $\mathbf{x}_1(s)$ and $\mathbf{x}_2(s)$. In this scenario, we observe that the design must accommodate for predictors with different degrees of the B-spline and number of knots, and their associated coefficient functions. The zero-degree B-spline expansion for $\mathbf{x}_1(s)$ results in simpler experimental runs, suitable for estimating the relatively simple coefficient function $\beta_1(s,t)$. In contrast, the second-degree B-spline expansion for $\mathbf{x}_2(s)$ provides the enough complexity to capture the intricacy of $\beta_2(s,t)$ which is expanded using first-degree B-splines.

Introducing multiple factors into a model increases the number of parameters that need to be estimated. In our experiments, we found that the coordinate exchange algorithm struggles to find an optimal design when the number of parameters is high. To address this issue, we can either expand the experiment by increasing the number of experimental runs or employ an alternative optimization algorithm.

In conclusion, the case of multiple functional predictors demonstrates that optimal experimental design can account for different complexities of the basis functions of each predictor and their associated coefficient functions. This example highlights that it is possible to accommodate the individual properties of each functional predictor and their associated coefficient functions in the design of experiments on the fully functional model.

\section{Proof of concept example}\label{sec:sim_ex}
\begin{figure}%
\centering
\subfloat[True coefficient]{\includegraphics[width=0.33\textwidth]{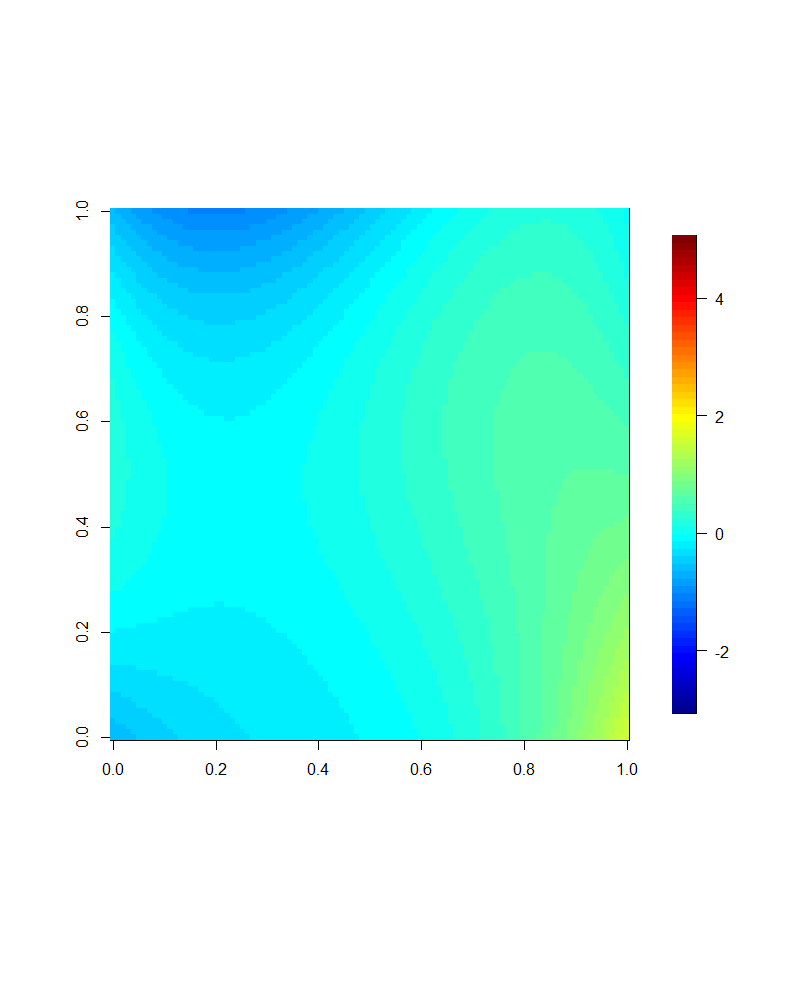}\label{fig:true coeff}}
\subfloat[Estimate from random design]{\includegraphics[width=0.33\textwidth]{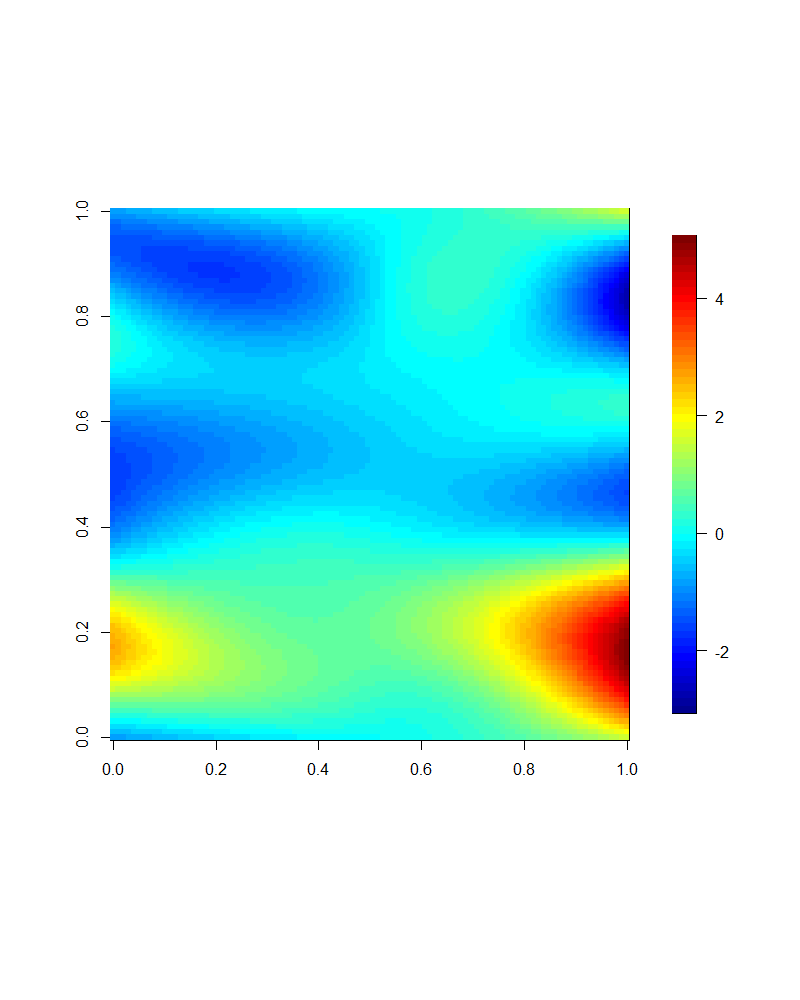}}
\subfloat[Estimate from A-optimal design]{\includegraphics[width=0.33\textwidth]{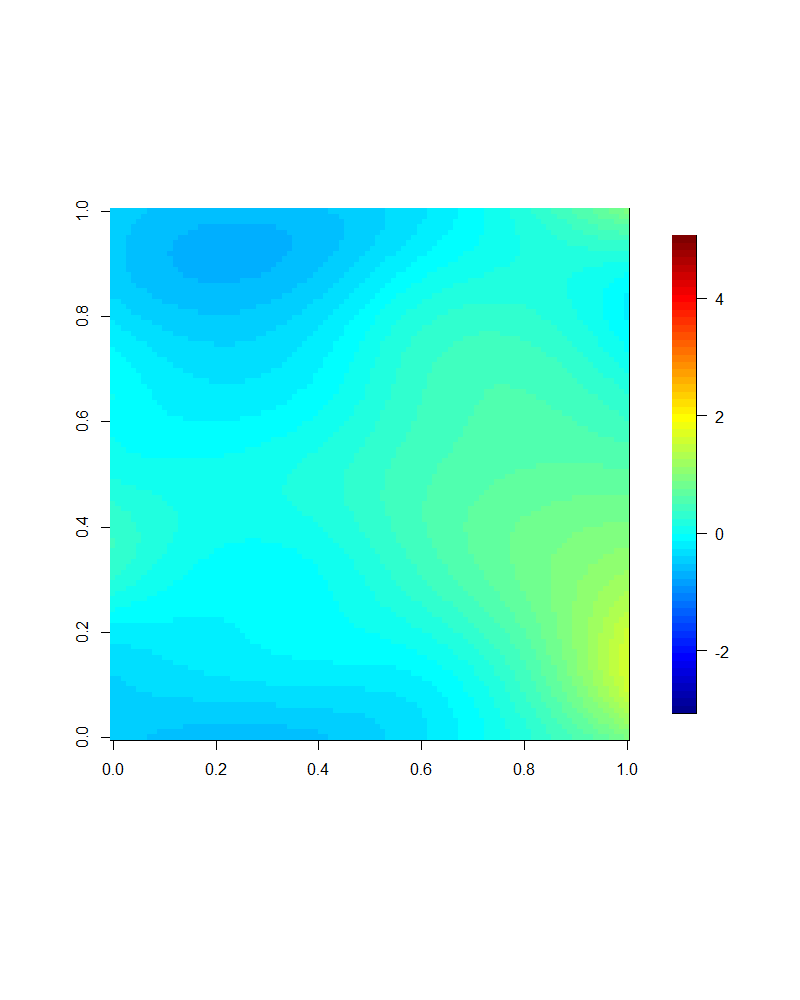}}
\hfill
\caption{True and estimated coefficients in the proof of concept example. The true coefficient (used to generate the responses) has been defined using 4 cubic B-splines in each direction. The estimators have been defined using $7$ Fourier basis functions in the response direction and $4$ second degree B-splines basis functions in the factor direction, and the factor is represented with $19$ first degree B-splines basis functions.}
\label{fig:proof_of_concept_results}
\end{figure}

In this section we present an illustrative example of the estimation of a function-on-function linear model using either a random design or the A-optimal design we derived. For simplicity, we consider the case of a single functional factor using the functional coefficient illustrated in Figure \ref{fig:true coeff}, and we use simulated data for the comparison. The true two-dimensional functional coefficient has been defined using separable basis functions, with cubic polynomials in both directions. The functional responses are then simulated from the model \eqref{model1mat2}, with the error term given by a Gaussian process with a radial basis function kernel with bandwidth parameter $\sigma^2=1e-04$ and variance $\gamma=0.005$ and represented with $81$ Fourier basis functions. These parameters have been chosen to generate functional responses with visually the smoothness and dispersion that are typical in functional data problems.

To obtain the A-optimal design, we need to decide the basis functions used to represent the functional factor and the basis functions to be used for the estimator of the coefficient in the factor direction (the basis functions $\{\eta_k(s), 1 \le k\le K\}$). Both these choices should be made by the experimenter, based on one side on the practical limitations and costs in the implementation of the functional factor and on the other side on the expected complexity of the functional coefficients. For this example, we are going to use one of the combination of basis functions we illustrated in the previous section, namely a first degree B-spline basis with $\mathcal{K}_X = 19$ for the functional factor and a second degree B-splines basis with $\mathcal{K}_\beta = 3$ for the functional coefficient. The corresponding A-optimal design can be seen in Figure  \ref{fig:scen-2.1-a}. As a comparison, we are going to use a random design where the coefficients of the basis functions for the functional factors are randomly sampled from a uniform distribution between $-1$ and $1$. 

To estimate the parameters of the model, we also need to choose basis functions for the other direction of the functional coefficients (i.e., $\{\theta_l(t), 1 \le l \le L\}$). This choice does not impact on the optimal design, as long as the chosen basis functions are orthonormal. For simplicity, we are going to use here a set of $7$ Fourier basis functions. We should note that the representation chosen for the coefficient does not match the one used for the true coefficient when simulating the data. This is intentional, since in a real scenario we do not know the functional form of the coefficient. The set of basis function on the response direction is also not the same that is used to represent the response curves in simulation, since in a real case the latter will depend on the amount of smoothing that is appropriate for the response curves. 

\begin{figure}%
\centering
\subfloat[Random design]{\includegraphics[width=0.5\textwidth]{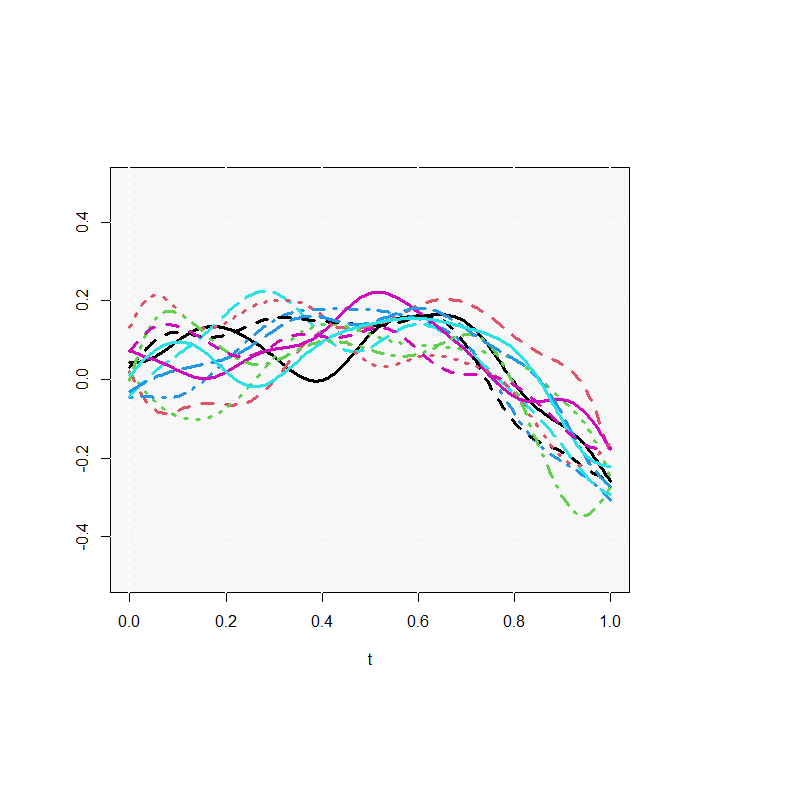}}
\subfloat[A-optimal design]{\includegraphics[width=0.5\textwidth]{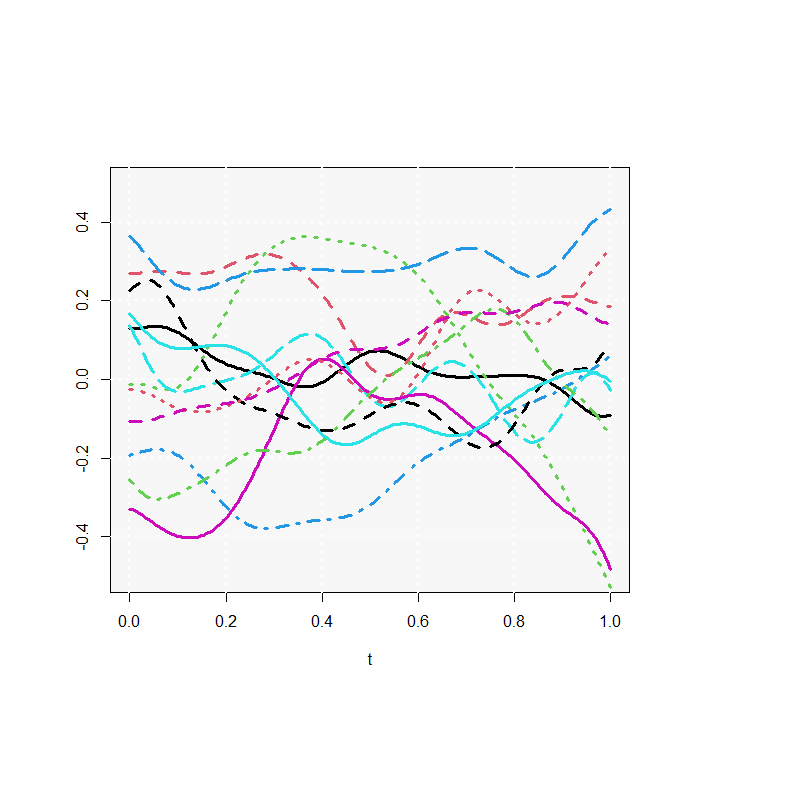}}
\caption{Response curves simulated from a function-on-function linear model with a single factor and the functional coefficient illustrated in Figure \ref{fig:true coeff}, using either a random design or the A-optimal design for the functional factor.}
\label{fig:proof_of_concept_responses}
\end{figure}

The simulated responses for the random design and the A-optimal design can be seen in Figure \ref{fig:proof_of_concept_responses}. They demonstrate the expected experimental results using the two different experimental designs. It can be seen that the random design results in response curves that are more concentrated, while the responses from the A-optimal design better explore the space of functional responses.

The two functional datasets can then be analysed using the R package \texttt{fda} \cite{fdap}. The corresponding estimated functional coefficients obtained using the \texttt{linmod} function can be seen in Figure \ref{fig:proof_of_concept_results}. Visually, the estimate from the A-optimal design is much closer to the true coefficient than the one obtained from a random design.

\section{Conclusions}\label{sec:concl}
This work presents new results on optimal experimental design for estimating the functional coefficient of a linear model with multiple dynamic factors and where also the response is functional. 
The first part of the paper develops inference results and extends optimality criteria for selecting designs of experiments, proving that both A-optimal and D-optimality do not depend on the covariance of the error process. 

The second part of the paper is dedicated to illustrating the proposed methodology in concrete examples.
In order to construct an experimental design in this context, an experimenter should set in advance the set of basis functions (e.g. type, order, dimension, etc) to be used for the predictors, and the set of basis functions to expand the functional coefficients in the direction of the predictors. As we have shown, the functional form of the optimal design is affected by this choice.

 The function-on-function model considered in this paper could be straightforwardly extended by including also
non-functional factors and concurrent functional interactions. The extension to time-delayed interactions is more challenging, since the model includes functional coefficients in higher dimensions. This will be scope for future work.
Moreover, in some cases it can be of interest to represent the functional coefficients using a higher number of basis functions and this can make the model not identifiable with the available number of runs. In this case, we need to include a penalization in the estimator, usually imposing a penalty on the derivatives of the functional coefficient. A further development will be to derive the expression of the optimal designs for this different estimator.

\backmatter

\bmhead{Supplementary information}

An exhaustive database of the experimental designs generated in this paper is available at https://shorturl.at/kgNzP. The dataset consists of 576 records, each representing one experimental design with 12 columns detailing the settings and results. The columns include the design ID (id); constants such as the number of random starts of the coordinate exchange algorithm (epochs) and the number of experimental runs (Runs); choices made for the functional factor—basis type (X family), degree (X degree), and number of breakpoints (X breaks); and similar choices for the functional coefficient (B family, B degree, B breaks). The results comprise the design coefficients of the basis expansion of the functional factor, the A-optimality criterion value, and the figure of the design. If a design cannot be estimated for specific settings, the design column will have the value "NONE," the criterion column will be null, and the figure column will contain no image.

\bmhead{Acknowledgements}
This work was supported by the Engineering and Physical Sciences Research Council, UK, [grant number EP/T021624/1, {\it Multi-objective optimal design of experiments}].
The second author's research has been supported by the EPSRC project 2607359.


\bibliography{f-bibliography}

\end{document}